\documentclass[aps,prl,twocolumn,showpacs,superscriptaddress]{revtex4}

\usepackage{graphicx}
\usepackage{amsmath}
\usepackage{amssymb}
\input{epsf}
\begin{document}

\title{Threshold behavior of bosonic two-dimensional few-body systems}

\author{D. Blume}
\affiliation{Department of Physics and Astronomy, Washington State University,
  Pullman, Washington 99164-2814}

\date{\today}

\begin{abstract}
Bosonic
two-dimensional self-bound clusters consisting of $N$ atoms 
interacting through additive van der Waals 
potentials become unbound at a critical
mass $m^{(N)}_*$; $m^{(N)}_*$ has been predicted to be independent
of the size of the system.
Furthermore, it has been predicted that the ground state energy $E_N$
of the $N$-atom system varies
exponentially as the atomic mass approaches $m_*$.
This paper reports accurate numerical many-body
calculations that 
allow these predictions to be tested. 
We confirm the existence of a universal critical mass $m_*$, and show that
the near-threshold behavior can only be described properly if a
previously neglected term is included.
We comment on the universality of the energy ratio $E_{N+1}/E_N$
near threshold.

\end{abstract}


\maketitle

Restricting the motion of particles to one or two dimensions can lead
to
properties that differ dramatically from those 
in three dimensions. 
The most prominent two-dimensional (2D) system
is the surface of bulk matter.
Another example are
one- or two-atom layer thin films, e.g.,
atomic or molecular hydrogen films~\cite{safo98,liu95}, grown on substrates. 
Neglecting the adatom-substrate interaction, many properties
of such systems can be 
understood within a truly 2D framework.
In addition to homogeneous 2D
systems, it is interesting to consider 
2D clusters~(see, e.g., Refs.~\cite{kris99,sars03}). What happens when a finite
number of atoms is restricted to 2D space? 
Inhomogeneous 2D systems can potentially be studied 
by placing a few atoms on the surface of
a substrate or by confining atoms by external potentials.
Effectively 2D
atom traps have been realized recently~\cite{goer01,rych04}; 
extension to optical
lattices with only a few atoms per lattice site is possible with
today's technology.
These systems are particularly interesting since Feshbach
resonances allow the interaction strengths to be tuned through application 
of magnetic fields. 

Bosonic 2D systems interacting through short-range two-body potentials 
that support one zero angular momentum bound state are predicted
to exhibit
intriguing universal, that is, model-independent, 
behaviors~\cite{bagc71,bruc76,bruc79,cabr79,lim80,tjon80,adhi88,adhi93,adhi95,jens04,hamm04,plat04}.
i) 
2D clusters with $N$ particles~\cite{bruc79,cabr79,tjon80}
are predicted to become unbound
when the mass $m$ reaches a critical value $m^{(N)}_*$~\cite{bagc71}.
This critical mass
is predicted to be universal, i.e., $m^{(N)}_*=m_*$~\cite{cabr79},
and to be the same for the corresponding
homogeneous system~\cite{bruc76,mill78}.
ii) For a given system size, the ground state energies $E_N$
near threshold are predicted to change 
exponentially as the atomic mass $m$ decreases~\cite{bagc71,lim80}. Similarly,
for a given atomic mass, the ground state energies near
threshold are predicted to change exponentially with varying system 
size~\cite{hamm04}. 
iii)
The ratio between the ground state energies 
of a 2D system 
with $N+1$ atoms and those of a system
with $N$ atoms reaches, in the limit of zero-range
interactions, a constant. 
This constant has been determined analytically for 
small 2D systems: $E_3^{\delta}/E_2^{\delta}=16.52$~\cite{bruc79,jens04} and 
$E_4^{\delta}/E_3^{\delta}=11.94$~\cite{plat04}. For large systems, 
the ratio $E_{N+1}^{\delta}/E_N^{\delta}$ has been predicted to approach 
8.57~\cite{hamm04}.
These predictions should apply to systems interacting through
short-range potentials if $|E_N| \ll \hbar^2/(m r_e^2)$,
where $r_e$ 
denotes the 
maximum of the characteristic length of the two-body potential
[in our case, the van der Waals length
$r_{\rm{vdW}}$, $r_{\rm{vdW}}= (mC_6/\hbar^2)^{1/4}$,
where 
$C_6$ denotes the leading
van der Waals coefficient] and the absolute value
of the effective range $r_{\rm{eff}}$.

This paper reports the energetics of
self-bound inhomogeneous 2D systems near threshold, and tests
under which conditions predictions i) to iii) apply. To this end
we perform numerical many-body calculations,
which supersede earlier variational 
calculations~\cite{cabr79,lim80}
and
cover a wider 
parameter range.
For $N=2$ and $3$ we determine the ground state energy 
using basis set expansion-type calculations. 
Since these techniques become untractable for larger systems,
we resort to Monte Carlo techniques for $N>3$.
We show that a proper description of the
ground state energies for 
small clusters with $N=2-5$ atoms
near threshold requires,
in addition to a term proportional to $(m-m^{(N)}_*)$,
a term proportional to $(m-m^{(N)}_*)^2$.
For larger droplets with
$N=6$ and 7 atoms we only include the linear term;
our diffusion Monte Carlo (DMC) energies do not allow 
the term proportional to $(m-m^{(N)}_*)^2$ to be determined. 
We speculate that DMC energies covering a larger mass range (see below)
would show that near threshold the quadratic term
is non-negligible
for systems with $N \ge 6$.
Our study confirms the existence of a universal
critical mass $m_*$~\cite{cabr79}.
This critical mass depends on the two-body 
potential but appears to be universal
for bosonic $N$-particle systems interacting additively
through a given two-body potential.
Finally, we investigate the ratio between the total ground state energy of 2D
systems with $N+1$ and $N$ particles. Our results are consistent with
predictions based on 
zero-range treatments~\cite{bruc79,jens04,plat04,hamm04} 
but do not conclusively confirm them for $N>3$. 
We comment on the relevance of the
energy scale associated with the effective range.

Consider the Hamiltonian $H$ for 
$N$ particles with mass $m$,
\begin{eqnarray}
	H = -\frac{\hbar^2}{2m} \sum_{i=1}^N\nabla_i^2 + \sum_{i<j}^N V(r_{ij}),
\end{eqnarray}
where $\nabla_i^2$ denotes the 2D Laplace operator of the $i$th
particle,
$r_{ij}$ the distance between atom $i$ and $j$, and $V$ the 
atom-atom potential. 
We use the realistic helium-helium potential without
retardation developed by
Korona {\em{et al.}}~\cite{koro97} (in the following
referred to as KORONA potential) as well as
a Lenard-Jones (LJ) potential with parameters
$\epsilon$ and $\sigma$ chosen to approximate the helium dimer 
interaction, i.e., $\epsilon=10.22$~K 
and $\sigma=2.56$~$\AA$~\cite{footnote3}. 
Both these two-body potentials support a single two-body
bound state with vanishing
angular momentum.
In the following,
we calculate the lowest energy with vanishing angular momentum
of the many-body system. We first treat ``true'' 
bosonic helium systems in 2D, i.e., we choose $m=m_{\rm{He}}$
(here and in the following, He denotes the isotope $^4$He). 
To explore 
the threshold regime, we then successively reduce the atomic mass 
while leaving the interaction potential unchanged. 

\begin{figure}
\centerline{\epsfxsize=3.0in\epsfbox{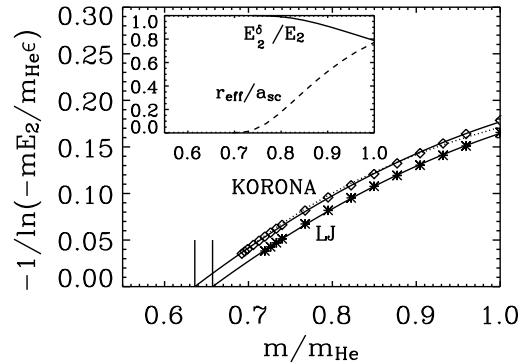}}
\vspace*{-1.3in}
\caption{
Scaled ground state energies $-1/\ln(m|E_2|/m_{\rm{He}}\epsilon)$ for
two 2D particles interacting through the KORONA potential 
(diamonds)
and through a simple LJ potential
(asterisks)
as a function of $m/m_{\rm{He}}$.
Solid lines show fits, which treat $m^{(2)}_*$, $a_1^{(2)}$ 
and $a_2^{(2)}$ as parameters.
Vertical solid lines 
indicate the mass ratio $m^{(2)}_*/m_{\rm{He}}$ at
which the fit predicts the dimer to become unbound.
The dotted line shows $E_2^{\delta}$, Eq.~(\protect\ref{eq_energyzero}),
for the dimer interacting through the KORONA potential.
The inset shows the ratios between the energies $E^{\delta}_2$
and $E_2$ (solid line), and between $r_{\rm{eff}}$ and $a_{sc}$ 
(dashed line) for two
particles interacting through the KORONA potential
as a function of 
$m/m_{\rm{He}}$. } \label{fig1}
\end{figure}

To start with, we solve the scaled 
radial Schr\"odinger equation 
for two atoms
by diagonalizing the
Hamiltonian using B-splines. For each mass,
the adaptive grid and box-size are optimized~\cite{footnote4}.
When we vary the atomic mass from $m/m_{\rm{He}}=1$  to 0.692,
the two-body binding energy $E_2$
for the KORONA potential
changes by nearly ten orders of magnitude from 
$E_2=-0.041$~K 
to $E_2=-8.6 \times 10^{-12}$~K.
The total ground state energies $E_N$ for a given 
$N$ near threshold, scaled by the atom mass $m$, are predicted to vary
exponentially as a function of $m$~\cite{bagc71},
\begin{eqnarray}
	mE_N \propto
	\exp \left\{- \left[ \sum_{i=1}^{\infty} a_i^{(N)}
	  \left
	  (\frac{m}{m_{\rm{He}}}- \frac{m^{(N)}_*}{m_{\rm{He}}} \right)^i \right]^{-1} \right \}.
	\label{eq_energyexp}
\end{eqnarray}
Here, $m^{(N)}_*$ denotes the critical mass at which the 
$N$-body system becomes unbound,
and $a_i^{(N)}$ parameters specific to the
$N$ particle system.
The mass $m$ is directly proportional to the coupling strength
$K$ and inversely proportional to the quantum parameter 
$\eta$, which have been used previously
to characterize LJ systems (see~\cite{footnote3}).
The functional form in Eq.~(\ref{eq_energyexp})
has been derived by Taylor-expanding the logarithmic derivative of
the bound state wave function for small total ground state energies
about $(m^{(N)}_*-m)$~\cite{bagc71}. Previous treatments neglected terms
proportional to $(m^{(N)}_*-m)^i$ with $i>1$. 
To test whether this is justified,
symbols in Fig.~\ref{fig1} show the 
quantity $-1/\ln(m|E_2|/ m_{\rm{He}}\epsilon)$, where $\epsilon$ denotes
the well depth
of the KORONA and LJ potential,
respectively,
as a function of $m/m_{\rm{He}}$
($\epsilon=11.06$~K 
for the KORONA potential and 
$10.22$~K for the LJ potential).
Diamonds show the ground state energies for the dimer interacting through
the
KORONA potential, and asterisks
those for the dimer interacting through 
the
LJ potential.
Figure~\ref{fig1}
shows clear deviations from a linear behavior,
indicating that the term proportional to 
$(m - m^{(2)}_*)^2$ cannot be neglected for $N=2$.
Earlier studies~\cite{cabr79,lim80,tjon80} 
did not see deviations from the linear behavior
possibly because
the total ground state energies were
i) varied over a smaller range, and
ii) determined variationally.
The qualitative behavior of the energies calculated using the
KORONA and the LJ potential is similar,
which implies that the non-linear 
behavior cannot be attributed to the difference in the long-range
parametrization of the two two-body potentials 
[the LJ potential is for large $r$ proportional
to $r^{-6}$, while the KORONA potential
contains additional terms proportional to $r^{-j}$, where
$j=8-16$ ($j$ even)].

Solid lines in Fig.~\ref{fig1}
show fits
of our scaled numerical two-body ground state energies 
to Eq.~(\ref{eq_energyexp}),
treating $m^{(2)}_*$, $a_1^{(2)}$ and $a_2^{(2)}$ as fitting parameters
and setting $a_i^{(2)}$ with $i>2$ to zero.
The fits 
predict that the two-body system interacting through the
KORONA potential becomes
unbound at $m^{(2)}_*=0.636(3)m_{\rm{He}}$ and that interacting through the
LJ potential
at $m^{(2)}_*=0.657(3)m_{\rm{He}}$. 
The numbers in brackets denote
the uncertainties of the fit, which are obtained by
including a varying number of data points in the fit.
Table~\ref{tab1} lists the fitting parameters 
$m_*^{(2)}$, $a_1^{(2)}$ and $a_2^{(2)}$ and their uncertainties.
The critical masses $m_*^{(2)}$ for both interaction potentials
are indicated in Fig.~\ref{fig1}
by vertical solid lines.

It is interesting to ask 
how well effective range theory describes
the near-threshold behavior
of the 2D dimer.
For a zero-range potential,
the ground state energy $E_2^{\delta}$ is determined by the
2D scattering length $a_{sc}$~(see, e.g., Ref.~\cite{jens04}),
\begin{eqnarray}
	E_2^{\delta} = -\frac{\hbar^2}{m a_{sc}^2} \; 4 \exp(-2 C),
	\label{eq_energyzero}
\end{eqnarray}
where $C$ denotes Euler's constant, $C=0.5772$.
Our definition of the scattering length $a_{sc}$ follows that adopted by
Verhaar {\em{et al.}}~\cite{verh84}, that is, the scattering wave function
goes through zero at $r=a_{sc}$.
In the following we restrict ourselves to dimers
interacting through the KORONA potential.
The scattering length $a_{sc}$, which we determine
numerically, varies from 77.8~\AA~
for the 2D helium dimer to about 
$5.7 \times 10^6$~\AA~
for the most weakly-bound dimer considered with $m = 0.692m_{\rm{He}}$. 
A dotted line in Fig.~\ref{fig1} shows
$E_2^{\delta}$, Eq.~(\ref{eq_energyzero}), using scaled dimensionless units.
Near threshold, $E_2^{\delta}$ nearly coincides with the 
numerically determined energy $E_2$ (diamonds).
At larger $m$, however, discrepancies are visible.
To quantify these discrepancies, a solid line in the inset of Fig.~\ref{fig1}
shows
the energy ratio $E_2^{\delta}/E_2$,
which varies from essentially 1 to $0.8$
for the atomic masses considered. 
Near threshold, the ground state energy $E_2$ can be described to
a very good approximation through a single atomic physics parameter, i.e.,
the two-body scattering length $a_{sc}$;
the importance of effective range corrections, however, 
increases with increasing mass.

To account for a non-vanishing effective range, the right hand side
of Eq.~(\ref{eq_energyzero}) has to be multiplied
by		
$\exp ( |E_2| \, m r_{\rm{eff}}^2/(4 \hbar^2)  )$~\cite{verh84,hamm04a}.
The effective range $r_{\rm{eff}}$,
\begin{eqnarray}
	r_{\rm{eff}} = 
2\sqrt{ 
\frac{ \hbar^2}{ |E_2|  m} 
\ln \left( |E_2| \frac{m a_{sc}^2}{\hbar^2} \frac{\exp(2C)}{4} \right),
	\label{eq_zerorange3}
}
\end{eqnarray}
can hence be evaluated
if the scattering length $a_{sc}$ and the binding energy $E_2$ are known.
For the dimers under study, $r_{\rm{eff}}$ changes from $60$~\AA~for
$m/m_{\rm{He}}=1$  to  $4900$~\AA~for
$m=0.696m_{\rm{He}}$. 
For smaller masses the numerical accuracy
of $E_2$ and $a_{sc}$ is not sufficient to reliably
determine $r_{\rm{eff}}$ from Eq.~(\ref{eq_zerorange3}).
A dashed line in the inset of Fig.~\ref{fig1} shows the ratio between
$r_{\rm{eff}}$ and $a_{sc}$.
Since the effective range $r_{\rm{eff}}$ increases with decreasing atomic
mass, the energy scale associated with the effective range $r_{\rm{eff}}$,
$\hbar^2/(m r_{\rm{eff}}^2)$~\cite{footnote1},
decreases with decreasing $m$ (see below).

We now turn to the study of 2D trimers interacting additively
through the KORONA potential. Since the potential energy depends on
relative coordinates only, we can separate off the center of mass motion.
Restricting ourselves to states with vanishing total angular momentum
reduces the number of degrees of freedom to three.
The 2D Schr\"odinger equation can then be rewritten in terms of three
hyperspherical coordinates. Here, we employ Whitten-Smith's democratic 
coordinates $R$, $\vartheta$ and $\varphi$~\cite{whit68}.
We determine the solution to the
Schr\"odinger equation by first calculating a set of 
angular-dependent channel functions and then solving a set of coupled 
hyperradial equations. 
Since our
B-spline implementation closely follows that used in Ref.~\cite{blum00a} for
3D trimers, we do
not discuss it in detail here. 
We note, however, that the 
grandangular momentum operator in 2D differs from that in 
3D~(see, e.g., Ref.~\cite{john83}) and that
the range of the angular coordinate $\vartheta$ changes
from $[0,\pi/4]$ in 3D to $[-\pi/4,\pi/4]$ in 2D~\cite{john80,john83}.
We include up to twelve channels
in our calculations and estimate the uncertainty
of the ground state 
energy $E_3$, which, in addition to the number of channels included in
the expansion,  depends on the 
angular and radial grids employed and on the step size $\Delta R$
used to calculate the coupling matrix elements~\cite{blum00a}, to be
at the few percent level.

\begin{figure}
\centerline{\epsfxsize=3.0in\epsfbox{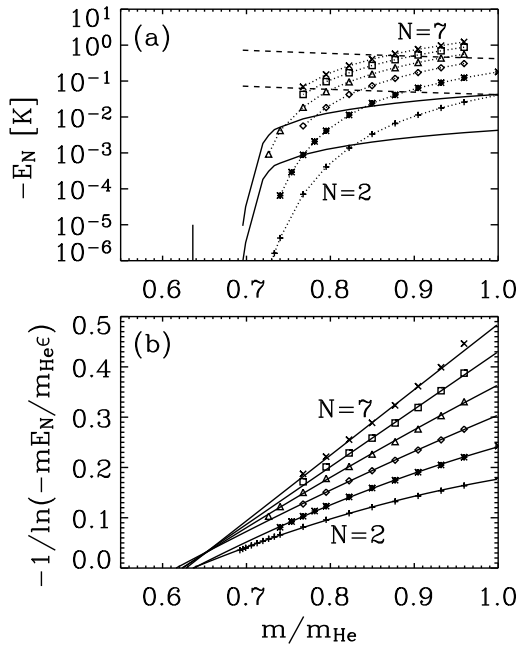}}
\caption{(a) Symbols show the total ground state energies $E_N$
for $N=2$ (pluses),
$N=3$ (asterisks),
$N=4$ (diamonds),
$N=5$ (triangles),
$N=6$ (squares), and
$N=7$ (crosses)
on a logarithmic scale 
as a function
of $m/m_{\rm{He}}$ for the KORONA potential.
Dotted lines connect data points for fixed $N$ to guide the eye.
The upper solid and dashed lines show $\hbar^2/(m r_e^2)$ 
for $r_e=r_{\rm{eff}}$ and $r_e=r_{\rm{vdW}}$, respectively.
The lower solid and dashed lines show $0.1\hbar^2/(m r_e^2)$ 
for $r_e=r_{\rm{eff}}$ and $r_e=r_{\rm{vdW}}$, respectively.
A vertical solid line indicates the 
mass ratio $m_*^{(2)}/m_{\rm{He}}= 0.636$,
at which our fits predict the 2D dimer to become unbound;
the critical masses for the larger systems nearly coincide with that for
the dimer.
(b) 
Scaled total ground state energies $-1/\ln(m |E_N|/m_{\rm{He}}\epsilon)$ 
as a function of
$m/m_{\rm{He}}$ using the same symbols as in (a).
Solid lines show fits 
(treating $m^{(N)}_*$, $a_1^{(N)}$ and $a_2^{(N)}$ as parameters for $N=2-5$, 
and treating $m^{(N)}_*$ and $a_1^{(N)}$ as parameters for $N=6-7$).} 
\label{fig2}
\end{figure}

Asterisks in Fig.~\ref{fig2}(a) show the
ground state energies $|E_3|$ on a log-scale as a function of
the scaled mass $m/m_{\rm{He}}$.
Since the system size increases with decreasing mass,
the calculations become more involved as the mass decreases.
For the masses considered, the total ground state energy varies
over nearly four orders of magnitude from $E_3 = -0.180$~K
for $m/m_{\rm{He}}=1$
to $E_3=-6.51 \times 10^{-5}$~K for $m/m_{\rm{He}}=0.740$.
To test the applicability of Eq.~(\ref{eq_energyexp}),
asterisks in
Fig.~\ref{fig2}(b) show the scaled gound state energies 
$-\ln(m |E_3|/m_{\rm{He}} \epsilon)$ as a function of the
scaled mass $m/m_{\rm{He}}$. 
As in the $N=2$ case, the scaled energies for $N=3$
clearly show deviations from
linear behavior. 
To determine the critical mass for the trimer system, we fit
our data to Eq.~(\ref{eq_energyexp}) treating 
$m_*^{(3)}$, $a_1^{(3)}$ and $a_2^{(3)}$ as fitting parameters.
Table~\ref{tab1} summarizes the result of the fit, which is 
shown by a solid line in Fig.~\ref{fig2}(b).
The uncertainties of the fit are, 
as in the dimer case, determined by including 
a varying number of data points in the fit.
Notably, the predicted critical mass $m_*^{(3)}$ for the trimer 
interacting additively through the KORONA potential
nearly coincides with the critical mass $m_*^{(2)}$ for the 
dimer interacting through the KORONA potential.

We now consider 2D systems  with up to $N=7$ particles
interacting additively through the KORONA potential. For these larger
systems, basis set expansion-type techniques become computationally
unfeasible. We thus 
solve the 2D many-body Schr\"odinger equation by alternative means using
the essentially exact DMC
technique with importance sampling~\cite{hamm94}. The DMC technique
allows ground state energies and structural properties
to be determined.
Since the numerical solution of the time-independent
Schr\"odinger equation is based on a stochastic process,
expectation values can only be determined within a statistical uncertainty.
This statistical uncertainty can be reduced by increasing the
computational efforts.
Our DMC energies for the 2D helium trimer and tetramer agree
with those calculated by Vranje\u{s} and Kili\'c~\cite{vran02}.
Furthermore, our DMC energies for the trimer 
agree to within the statistical uncertainty
with those calculated by the hyperspherical B-spline treatment (see above).
In the following, we report the ground state 
energies $E_N$ for $N>3$ calculated by the DMC method as a function of
the atomic mass.
As we approach the threshold regime,
the DMC calculations become more difficult
since the kinetic and potential energy nearly cancel.

Symbols in Fig.~\ref{fig2}
show the DMC ground state energies $E_N$ as a function of the mass ratio 
$m/m_{\rm{He}}$ for $N=4-7$.
Panel~(a) shows the energies on a logarithmic scale
(to guide the eye dotted lines connect data points for the same $N$).
To investigate how well the functional form given in Eq.~(\ref{eq_energyexp})
applies to systems with $N > 3$,
panel~(b) shows the scaled
DMC ground state 
energies $-1/\ln(m |E_N|/ m_{\rm{He}} \epsilon)$.
Statistical uncertainties of the DMC energies (not shown) are smaller
than the symbol size.
Deviations from a linear behavior are, although less pronounced than
for the dimer and trimer, 
visible for $N=4$ and $5$; consequently, we
fit our DMC energies to Eq.~(\ref{eq_energyexp}),
treating $m^{(N)}_*$, $a_1^{(N)}$ and $a_2^{(N)}$ as fitting parameters.
For the clusters with $N=6$ and $7$, 
our scaled DMC energies depend to a good approximation linearly on the mass;
hence we use only two fitting parameters, $m_*^{(N)}$ and $a_1^{(N)}$.
The fitting parameters are summarized in Table~\ref{tab1}, and the fits
are shown by
solid lines in Fig.~\ref{fig2}(b).
The statistical uncertainty of the fits is, as in the dimer and trimer case,
estimated by including a varying number of data points in the fit.
We speculate that, if we were 
able to obtain accurate DMC energies closer to threshold, 
non-linear behavior of the scaled ground state energies 
for 
$N \ge 6$ 
would be, similarly as for the dimer and trimer, visible.
The fits for $N=4-7$ predict critical
masses $m^{(N)}_*/m_{\rm{He}}$ between $0.62$ and
$0.63$, which agree within their uncertainties
with those
predicted for $N=2$ and $3$.
Our analysis thus
confirms that the critical mass $m^{(N)}_*$ at which 2D systems,
interacting additively through a given two-body potential,
become unbound is the same for all system sizes~\cite{cabr79,lim80,tjon80}.

\begin{figure}[]
\centerline{\epsfxsize=3.0in\epsfbox{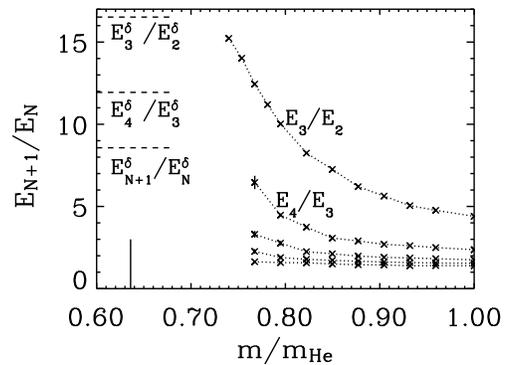}}
\vspace*{-1.3in}
\caption{
Energy ratios $E_{N+1}/E_N$ as a function of 
$m/m_{\rm{He}}$ for 
$N=2$ (uppermost curve) through
$N=6$ (lowermost curve) calculated using the KORONA potential.
Errorbars reflect the statistical uncertainties of our DMC energies
$E_N$ for $N \ge 4$.
Dashed horizontal lines on the left hand side indicate the
energy ratios
for zero-range interactions:
$E_3^{\delta}/E_2^{\delta}=16.52$,
$E_4^{\delta}/E_3^{\delta}=11.94$ and
$E_{N+1}^{\delta}/E_N^{\delta}=8.57$. 
A vertical solid line indicates the 
mass ratio $m_*^{(2)}/m_{\rm{He}}= 0.636$,
at which our fits predict the 2D dimer to become unbound.
The critical masses for larger systems nearly coincide 
with that for the dimer.
} \label{fig3}
\end{figure}

To further investigate the near-threshold behavior 
for $N>2$,
symbols in Fig.~\ref{fig3} show the ratio between the ground state
energies for systems with $N+1$ and $N$ atoms as a function of
$m/m_{\rm{He}}$. To guide the eye, dotted lines connect
data points for $E_{N+1}/E_N$ with the same $N+1$ and $N$ but
different $m$.
Errorbars,
which increase with decreasing mass,
reflect the statistical uncertainty of our DMC
energies for $N \ge 4$.
At large $m$, the energy ratios approach a constant,
in agreement with earlier studies~\cite{tjon80}.
Since $|E_2|$ decreases with decreasing mass 
while the van der Waals length
$r_{\rm{vdW}}$ changes only little [see dashed lines in Fig.~\ref{fig2}(a)],
dimers with small $m$ effectively have a shorter range
than those with large $m$.
Accordingly, systems with small $m$ should be better described
by zero-range models than those with large $m$.
Horizontal dashed lines on the left hand side of Fig.~\ref{fig3} indicate the
energy ratios $E_3^{\delta}/E_2^{\delta}=16.52$~\cite{jens04},
$E_4^{\delta}/E_3^{\delta}=11.94$~\cite{plat04} and
$E_{N+1}^{\delta}/E_N^{\delta}=8.57$~\cite{hamm04} for
2D systems interacting through additive zero-range potentials.
While the energy ratio $E_3/E_2$  calculated for the helium-like
few-body systems with small $m$
is close to the value predicted by the zero-range treatment,
none of the energy ratios $E_{N+1}/E_N$ for $N \ge 3$
is.
Although the energy ratios for $N \ge 3$ 
increase with decreasing mass, it is not clear
whether they approach the values predicted for the zero-range model.

Recall that zero-range treatments should
be applicable when the energy scale associated with
$r_e$
is much larger than $|E_N|$,
where
$r_e$ is either given by the 
van der Waals length $r_{\rm{vdW}}$ or by the effective range $r_{\rm{eff}}$.
Solid and dashed lines in panel~(a) of Fig.~\ref{fig2} 
show the energy 
$\hbar^2/(m r_e^2)$
for $r_e = r_{\rm{eff}}$ (upper solid line)
and for $r_e = r_{\rm{vdW}}$ (upper dashed line), respectively.
Since $|E_N| \ll \hbar^2/(m r_e^2)$ for zero-range treatments to be valid, 
the lower solid and dashed lines show the quantity $0.1\,\hbar^2/(m r_e^2)$
for $r_e=r_{\rm{eff}}$ and $r_e=r_{\rm{vdW}}$, respectively.
Figure~\ref{fig2}(a) indicates that, of the systems considered,
only those with $N=2$ and $3$ have
total ground state energies $|E_N|$ that are
smaller than the lower solid and dashed lines.
Although the dimer near threshold is well described by a zero-range treatment
(see Fig.~\ref{fig1}), the corresponding systems with $N>3$ cannot
be properly described through simple contact interactions. 
This is due to the fact that the total ground state energies 
vary exponentially as a function of $N$ and that 
the effective range $r_{\rm{eff}}$ increases with decreasing $m$.
To check whether the latter is specific to the KORONA 
potential studied here, we also calculated
the effective range 
$r_{\rm{eff}}$ for the
LJ potential discussed above and for a realistic tritium-tritium
b$^3\Sigma_u^+$ potential.
For the three potentials considered, the effective
range behaves similarly
as a function of the two-body binding energy.
Our findings suggest that the regime where the $N$-body
total ground state energies $E_N$, $N \gg 2$, 
can be properly described by zero-range models
might, in general, be
hard to reach for 2D van der Waals
systems. 

\begin{table}
\begin{tabular}{l|llll} 
& $N$ & $m_*^{(N)}$ & $a_1^{(N)}$ & $a_2^{(N)}$  \\ \hline
LJ 	&2 	& 0.657(3) &	0.66(2) &	-0.53(6) \\ \hline
KORONA 	&2  	& 0.636(3) &	0.68(2) &	-0.55(6) \\
&3 &	0.637(8) &	0.86(5) &	-0.53(14) \\
&4 &	0.62(2) &	0.89(8)	&	-0.25(12) \\
&5 &	0.62(2) & 	1.06(10)	&	-0.26(20) \\
&6 &	0.62(3) &	1.14(10)	&	 \\
&7 &	0.63(2) &	1.30(7)	&	 
\end{tabular}
\caption{Fitting parameters 
for 2D clusters interacting through the Lenard-Jones potential
(for $N=2$ only) and through the KORONA potential ($N=2-7$), respectively.
We treat three fitting parameters for $N=2-5$, 
and two fitting parameters for $N=6-7$.
The numbers in brackets give the uncertainties of the fitting parameters,
which are obtained by including a varying number of data points in the fit.}
\label{tab1}
\end{table}

Finally we remark on 
an earlier variational study~\cite{kris99}, which 
concluded that artificial ``bosonic helium 3'' clusters in 2D
become bound for 
a minimum of about twelve atoms.
Although the figures in this paper contain energies for
$m \le m$($^3$He) for only a few system sizes, our 
calculations show that bosonic $^3$He clusters 
are self-bound for all $N$ considered, in agreement with Ref.~\cite{vran02}.
(We excluded some energies for small $m$ from the figures and 
the fits since their statistical uncertainties are very large.)
Furthermore, the existence of a universal critical mass
$m_*$ indicates
that ``bosonic helium 3'' 2D clusters, which are often studied
to complement the treatment of fermionic helium 3 systems, are bound for all
$N$.
Our calculations emphasize 
that great care has to be taken when variational calculations are used
to investigate the near-threshold regime (see also Ref.~\cite{sars03}).

Fruitful discussions with H.-W. Hammer and D.T. Son, which motivated this
work, and with C.H. Greene are gratefully acknowledged.
Acknowledgement is made to the Donors of The
Petroleum Research
Fund, administered by the American Chemical Society, 
and the NSF (grant ITR-0218643) for support of this research.


\begin{thebibliography}{10}

\bibitem{safo98}
A.~I. Safonov {\it et~al.}, Phys. Rev. Lett. {\bf 81},  4545  (1998).

\bibitem{liu95}
F.-C. Liu, Y.-M. Liu, and O.~E. Vilches, Phys. Rev. B {\bf 51},  2848  (1995).

\bibitem{kris99}
B. Krishnamachari and G.~V. Chester, Phys. Rev. B {\bf 59},  8852  (1999).

\bibitem{sars03}
A. Sarsa, J. Mur-Petit, A. Polls, and J. Navarro, Phys. Rev. B {\bf 68},
  224514  (2003).

\bibitem{goer01}
A. G\"orlitz {\it et~al.}, Phys. Rev. Lett. {\bf 87},  130402  (2001).

\bibitem{rych04}
D. Rychtarik {\it et~al.}, Phys. Rev. Lett. {\bf 92},  173003  (2004).

\bibitem{bagc71}
A. Bagchi, Phys. Rev. A {\bf 3},  1133  (1971).

\bibitem{bruc76}
L.~W. Bruch, Phys. Rev. B {\bf 13},  2873  (1976).

\bibitem{bruc79}
L.~W. Bruch and J.~A. Tjon, Phys. Rev. A {\bf 19},  425  (1979).

\bibitem{cabr79}
F. Cabral and L.~W. Bruch, J. Chem. Phys. {\bf 70},  4669  (1979).

\bibitem{lim80}
T.~K. Lim, S. Nakaichi, Y. Akaishi, and H. Tanaka, Phys. Rev. A {\bf 22},  28
  (1980).

\bibitem{tjon80}
J.~A. Tjon, Phys. Rev. A {\bf 21},  1334  (1980).

\bibitem{adhi88}
S.~K. Adhikari {\it et~al.}, Phys. Rev. A {\bf 37},  3666  (1988).

\bibitem{adhi93}
S.~K. Adhikari, A. Delfino, T. Frederico, and L. Tomio, Phys. Rev. A {\bf 47},
  1093  (1993).

\bibitem{adhi95}
S.~K. Adhikari, T. Frederico, and I.~D. Goldman, Phys. Rev. Lett. {\bf 74},
  487  (1995).

\bibitem{jens04}
A.~S. Jensen, K. Riisager, D.~V. Fedorov, and E. Garrido, Rev. Mod. Phys. {\bf
  76},  215  (2004).

\bibitem{hamm04}
H.-W. Hammer and D.~T. Son, {Phys. Rev. Lett.} {\bf 93},  250408  (2004).

\bibitem{plat04}
L. Platter, H.-W. Hammer, and U.-G. Meissner, {Few-Body Systems} {\bf 35},  169
   (2004).

\bibitem{mill78}
M.~D. Miller and L.~H. Nosanow, J. Low Temp. {\bf 32},  145  (1978).

\bibitem{koro97}
T. Korona {\it et~al.}, J. Chem. Phys. {\bf 106},  5109  (1997).

\bibitem{footnote3}
{The calculations are performed using (all digits used in the calculation are
  reported) $m_{{\rm{He}}}=7296.3$~m$_e$, $\epsilon=10.22$~K, $\sigma
  =2.56$~\AA~and $V_{LJ} = 4 \epsilon [ (r/\sigma)^{-12}-(r/\sigma)^{-6}]$.
  These parameters result in the dimensionless coupling constant $K=22.11$
  (rounded), where $K=4m \epsilon \sigma^2 / \hbar^2$, and the quantum constant
  $\eta=0.1809$ (rounded), where $\eta = \hbar^2/(m \epsilon \sigma)$.}

\bibitem{footnote4}
{ The calculations require the maximum $r$ value $r_{max}$ to be chosen
  properly; to obtain full convergence $r_{max}$ has to be roughly 100 times
  larger than the 2D two-body scattering length $a_{sc}$ (see text). To
  converge our two-body energies, we space the grid points along the radial
  coordinate $r$ using a $r^{1/n}$ grid, where $n$ denotes an integer, $n=1-8$
  (the larger $r_{max}$ the larger $n$). Convergence was tested by varying the
  number of grid points between 400 and 800.}

\bibitem{verh84}
B.~J. Verhaar, J.~P. H.~W. {van den Eijnde}, M.~A.~J. Voermans, and M.~M.~J.
  Schaffrath, J. Phys. A {\bf 17},  595  (1984).

\bibitem{hamm04a}
H.-W. Hammer, {private communication.}  

\bibitem{footnote1}
{In defining the energy scale associated with $r_e$ we drop the factor $4
  \exp(-2C)=1.2609$.}

\bibitem{whit68}
R.~C. Whitten and F.~T. Smith, J. Math. Phys. {\bf 9},  1103  (1968).

\bibitem{blum00a}
D. Blume, C.~H. Greene, and B.~D. Esry, J. Chem. Phys. {\bf 113},  2145
  (2000).

\bibitem{john83}
B.~R. Johnson, J. Chem. Phys. {\bf 79},  1916  (1983).

\bibitem{john80}
B.~R. Johnson, J. Chem. Phys. {\bf 73},  5051  (1980).

\bibitem{hamm94}
B.~L. Hammond, W.~A. {Lester, Jr.}, and P.~J. Reynolds, {\em Monte Carlo
  Methods in Ab Initio Quantum Chemistry} (World Scientific, Singapore,
  1994).

\bibitem{vran02}
L. Vranje\u{s} and S. Kili\'c, Phys. Rev A {\bf 65},  042506  (2002).

\end{thebibliography}

\end{document}